\documentstyle[11pt,newpasp,twoside,psfig]{article}

\index{supernova remnants!G347.3--0.5}
\index{supernova remnants!interactions with molecular clouds}
\index{supernova remnants!synchrotron emission}

\def\edcomment#1{\iffalse\marginpar{\raggedright\sl#1\/}\else\relax\fi}
\reversemarginpar

\begin{document}

\title{\textit{XMM-Newton} observation of SNR RX J1713.7--3946}
\author{\textsc{G.~Cassam-Chena\"{i}}, \textsc{A.~Decourchelle}, \textsc{J.~Ballet}, \textsc{J.-L.~Sauvageot}}
\affil{Service d'Astrophysique, CEA Saclay, 91191 Gif-sur-Yvette, France}
\author{\textsc{G.~Dubner}}
\affil{Instituto de Astronom\'{i}a y F\'{i}sica del Espacio, CC 67, Suc. 28, 1428, Buenos Aires, Argentina}

\begin{abstract}
We present the first results of the observations of the supernova remnant RX J1713.7--3946 (also G347.3--0.5)
obtained with the EPIC instrument on board the \textit{XMM-Newton} satellite.
We show a 5 pointings mosaiced image of the X-ray synchrotron emission.
We characterize this emission by mapping its spectral parameters 
(absorbing column density $N_{\mathrm{H}}$ and photon index $\Gamma$).
The synchrotron spectrum is flat at the shock and steep in the interior of the remnant.
$N_{\mathrm{H}}$ is well correlated with the X-ray brightness.
A strong $N_{\mathrm{H}}$ is found in the southwest rim of RX J1713.7--3946.
We suggest that the SNR is interacting with a H\textsc{i} region there.
\end{abstract}

\section{Introduction}\label{introduction}

RX J1713.7--3946 is a shell-type supernova remnant (SNR) located in the Galactic 
plane that was discovered with the \textit{ROSAT} all-sky 
survey (Pfeffermann \& Aschenbach 1996).
The observation of the northwestern shell of the SNR with the \textit{ASCA} 
satellite has shown the presence of pure non-thermal emission 
(Koyama et al. 1997).
Further observations with \textit{ASCA} of most of 
the remnant did not reveal traces of the thermal emission, 
being likely overwhelmed by the bright X-ray synchrotron emission 
(Slane et al. 1999). 

Both the distance and age of RX J1713.7--3946 are still not well known. 
Based on the X-ray measurement of the column density toward 
this source, Koyama et al. (1997) derived 
a distance of 1 kpc, while Wang et al. (1997) 
argued that it exploded in AD 393. 
Subsequently, Slane et al. (1999) proposed a larger distance of 6 kpc 
based on its probable association with three dense and massive molecular clouds.
Assuming a Sedov phase evolution, 
an age of a few $10^4$ years is derived for such a distance.

In the radio, the emission arises from faint filaments aligned with the 
X-ray shell of RX J1713.7--3946.
GeV and TeV $\gamma$-ray emissions were detected by \textit{EGRET}
to the northeast of the SNR (Hartman et al. 1999; Butt et al. 2001) and 
by \textit{CANGAROO} in the northwest (Muraishi et al. 
2000; Enomoto et al. 2002).
The association of these high energy emissions with RX J1713.7--3946
has been the subject of debate 
(Reimer \& Pohl 2002, Butt et al. 2002).

In this paper, we intend to give for the first time a detailed description of 
the X-ray emission of RX J1713.7--3946 with the 
\textit{XMM-Newton} observatory.
The high sensitivity of \textit{XMM-Newton} will allow us to carry out a spectral 
analysis at medium scale of the emission structures and 
then to produce for the first time a mapping of the spectral parameters 
of RX J1713.7--3946.

\section{Observations}\label{data red}
SNR RX J1713.7--3946 was observed with the three EPIC instruments 
(namely the MOS1, MOS2 and pn cameras) 
on board the \textit{XMM-Newton} satellite in the course of the AO-1 program.
The observations were performed in five distinct pointings, 
each of around 10 ks duration.
They correspond schematically to the center, northeast, 
northwest, southwest and southeast 
(hereafter CE, NE, NW, SW, SE, respectively) parts of RX J1713.7--3946.

\begin{figure}
\centering
\begin{tabular}{cc}
\psfig{file=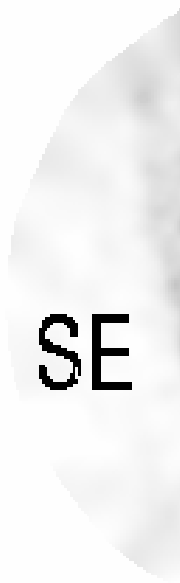,height=4.5cm} & \psfig{file=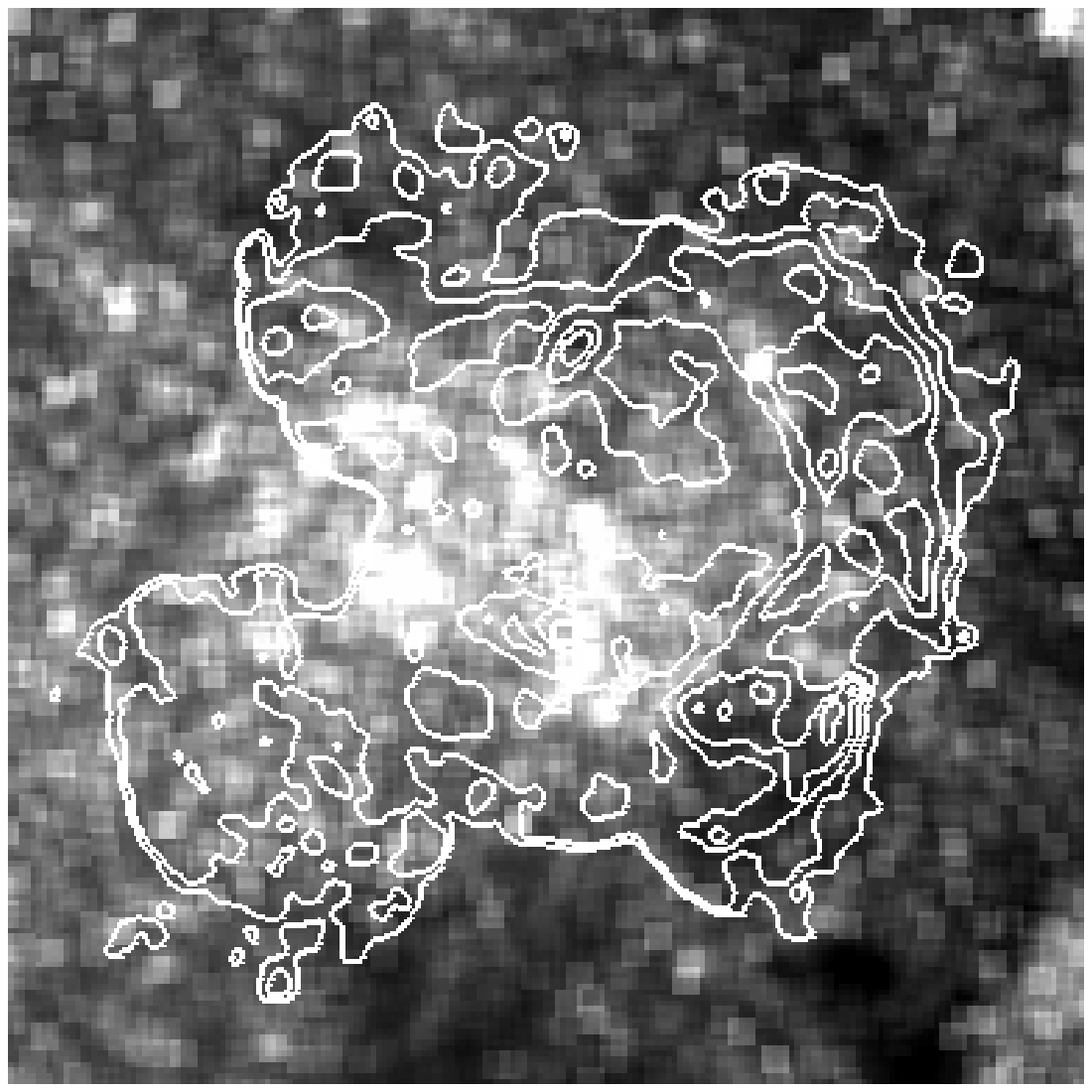,height=4.5cm} \\
\psfig{file=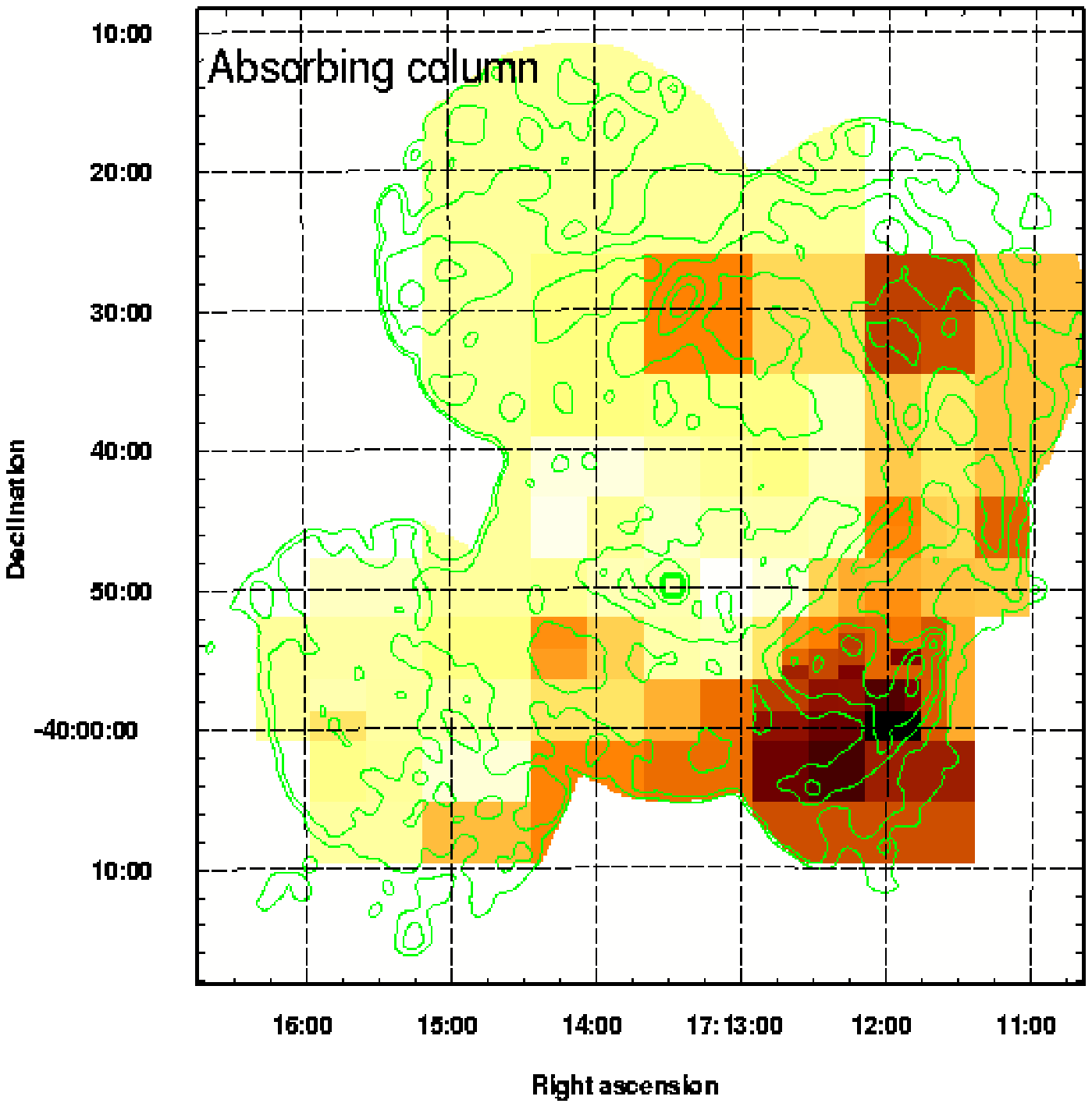,width=6cm} & \psfig{file=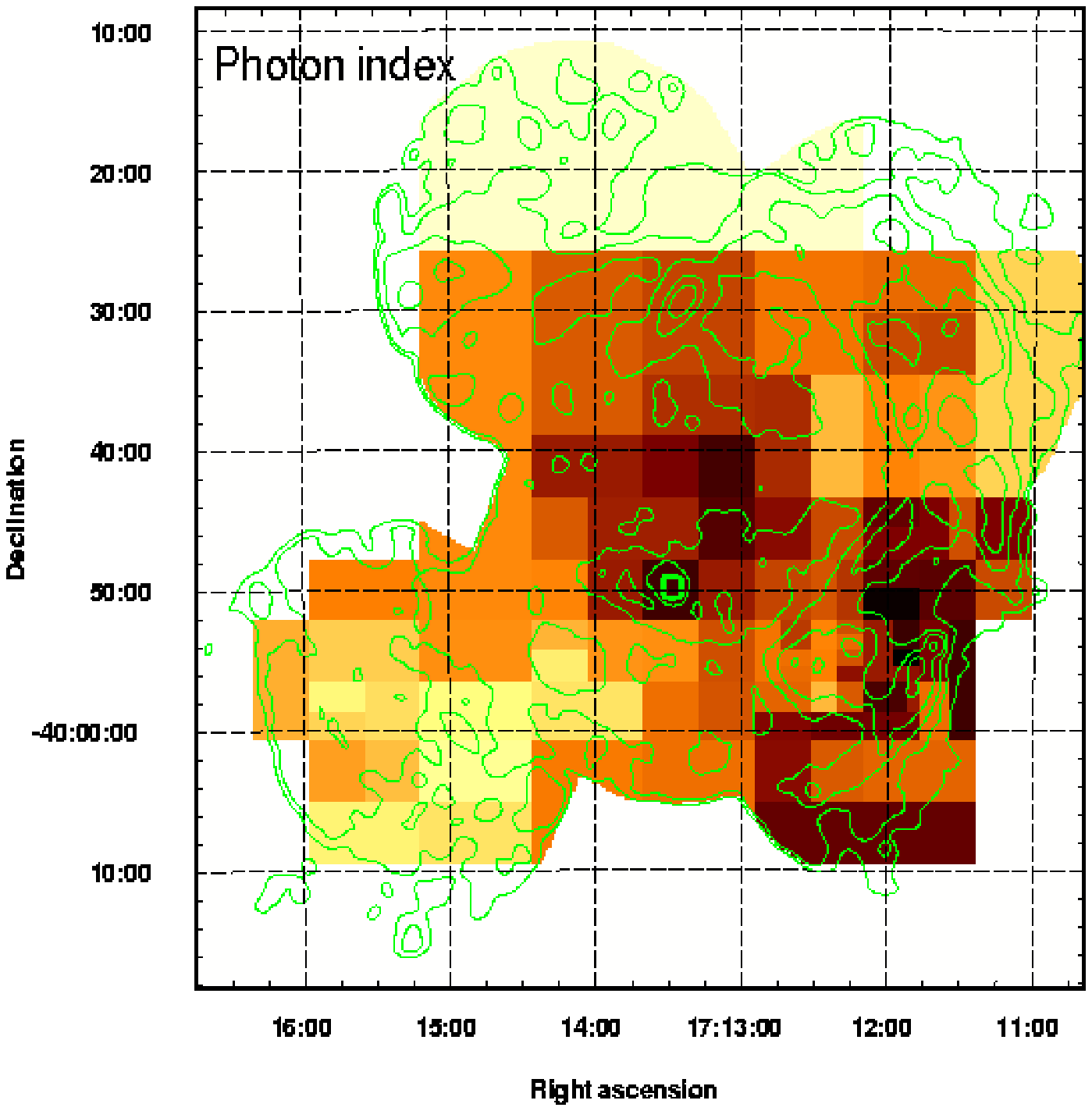,width=6cm} \\
\psfig{file=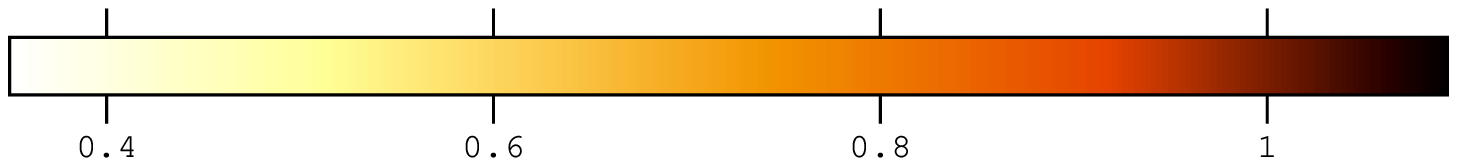,width=5cm,height=0.5cm} & \psfig{file=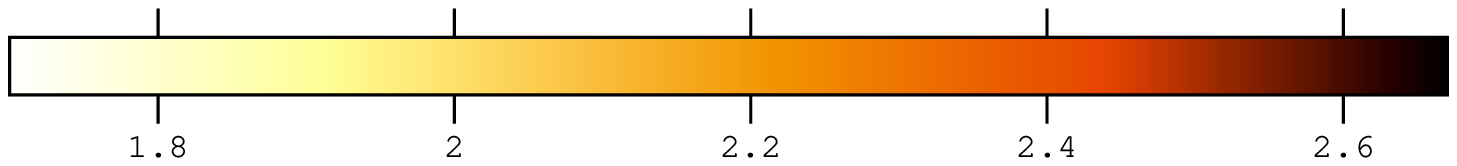,width=5cm,height=0.5cm} 
\end{tabular}
\caption{\textit{Top-left panel:} adaptively smoothed image of RX J1713.7--3946 in the 2-10 keV energy band.
\textit{Top-right panel:} Smoothed optical image (DSS in red color) overlaid with the 2-10 keV contours.
\textit{Bottom-left panel:} Absorbing column density map overlaid with the 2-10 keV contours.
\textit{Bottom-right panel:} Photon index map overlaid with the 2-10 keV contours.}
\label{fig1}
\end{figure}

\section{Spectro-imaging of SNR RX J1713.7--3946}\label{spectro-imaging}

\subsection{X-ray morphology}\label{x_ray_morpho}
Figure \ref{fig1} (top-left panel) shows the mosaiced image 
of RX J1713.7--3946 in the 2-10 keV energy band.
The 5 pointings do not permit to cover the entire remnant (notably in the East) and show
an unclassical X-ray synchrotron morphology.
The brightest regions found in the west are also the more structured.
At large scale, a kind of double shell stands out.
A first shell toward the interior is more or less circular 
whereas a second one, less easy to imagine, is split into a few arcs.

\subsection{Spatial and spectral characterizations of the synchrotron emission}\label{syn_emis}
To map and characterize the spectrum of the non-thermal emission, 
we create an adaptative spatial grid from which the X-ray spectra (MOS, pn) are extracted and
fitted with a simple power-law.
Each pixel of the grid has approximately $9000$ counts (MOS+pn).
The adjusted value of the spectral parameters (absorbing column density $N_{\mathrm{H}}$
and photon index $\Gamma$) is then attributed to the corresponding pixel grid to construct the map.

Figure \ref{fig1} (bottom panels) shows the $N_{\mathrm{H}}$ (left) and $\Gamma$ (right) maps 
obtained by the method described above.
The variations of absorbing column density are strong since $N_{\mathrm{H}}$ 
varies from $\sim 0.4 \: 10^{22}$ cm$^{-2}$ to $\sim 1.1 \: 10^{22}$ cm$^{-2}$.
The mean relative error on the absorbing column in each pixel grid is $8.5\%$ with a maximum value of $16\%$.
In the SE and CE, $N_{\mathrm{H}}$ is low with a value $\sim 0.4-0.5 \: 10^{22}$ cm$^{-2}$ 
whereas it is larger in the NW ($N_{\mathrm{H}} \sim 0.6-0.7 \: 10^{22}$ cm$^{-2}$) 
and SW ($N_{\mathrm{H}} \sim 0.8-1.1 \: 10^{22}$ cm$^{-2}$).
It is surprising to find a high value in the SW since no molecular clouds are found there.
This point is discussed in Sect. \ref{interaction_with_clouds}.

Figure \ref{fig1} (bottom-right panel) shows that the variations of photon index are important. 
$\Gamma$ varies from $\sim 1.8$ to $\sim 2.6$.
The mean relative error on $\Gamma$ in each pixel grid is $3.8\%$ with a maximum value of $4\%$.
The spectrum is steep in the faint CE region and flat at the shock particularly in the SE and NW,
except in a few places in the SW which are coincident with the faint X-ray emission.

\subsection{Interaction with clouds}\label{interaction_with_clouds}
Figure \ref{fig1} (bottom-left panel) shows that the absorbing column density
and the X-ray brightness are correlated. 
For instance, the $N_{\mathrm{H}}$ is large where the X-ray brightness is strong.
If the absorbing column was not related to the remnant, we would not expect any
link between the $N_{\mathrm{H}}$ and the X-ray brightness.
That we observe a correlation means that the increased absorption density somehow amplifies the X-ray brightness.
This can be interpreted as the result of the SNR interaction 
with some absorbing material in the brightest regions.

It is possible to estimate the density of the absorbing material 
if one assumes that the angular size $\theta_{\mathrm{abs}}$ of the absorbing material is the same
along the line-of-sight and in projection on the sky.
If it is so, the density of the absorbing matter is
\begin{equation}\label{n_abs}
n_{\mathrm{abs}} = 6.68 \; 10^{5} \; \left(\frac{D}{1 \; \mathrm{kpc}}\right)^{-1} 
\left(\frac{\theta_{\mathrm{abs}}}{1 \arcsec}\right)^{-1} 
\left(\frac{\Delta N_{\mathrm{H}}}{10^{22} \; \mathrm{cm}^{-2}}\right) \; \mathrm{cm}^{-3}
\end{equation}
where $D$ is the SNR distance and $\Delta N_{\mathrm{H}}$ is the variation of the 
absorbing column density at scale $\theta_{\mathrm{abs}}$.
Typically in the SW region, $\theta_{\mathrm{abs}} \sim 15\arcmin$ and 
$\Delta N_{\mathrm{H}} \simeq 0.4 \: 10^{22}$ cm$^{-2}$.
At 6 kpc, it yields $n_{\mathrm{abs}} \sim 50 \; \mathrm{cm}^{-3}$.

To check that the additional column density is real, we have looked at the map of integrated star light 
(Fig. \ref{fig1}, top-right panel)
which should somehow reflect $N_{\mathrm{H}}$ variations.
This test confirms a good correlation between our inferred column density and the optical brightness
which makes us confident in our $N_{\mathrm{H}}$ map.

The pending question is then to determine the nature of the matter interacting with SNR RX J1713.7--3946.
In the west side of the remnant, there is no evidence for an interaction with
molecular clouds since CO observations have shown that they are located in the north. 
The shock front of the remnant may have reached a stellar wind bubble shell.
This is a possible and already proposed scenario (Slane et al. 1999) 
due to the low density medium in which the remnant seems to evolve.
However, the regions of low integrated star light extend much further 
than the remnant's extent (Fig. \ref{fig1}, top-right panel)
whereas a stellar wind shell would be relatively thin.
So it is also possible that a preexisting H\textsc{i} region is interacting with SNR RX J1713.7--3946.
This point is supported by the fact that 
the density that we derive from Eq. (\ref{n_abs}) is typical of 
H\textsc{i} densities in the ISM. 
Recent H\textsc{i} observations at high resolution around RX J1713.7--3946 (Koo et al. 2004) also
indicate an excess in that direction.

\acknowledgments
G.C.-C. would like to thank cordially Jocelyn Bell Burnel and Joseph H. Taylor for their encouragements.

\end{document}